\documentclass[conference, 10pt]{IEEEtran}
\IEEEoverridecommandlockouts 
\usepackage[]{graphicx}
\graphicspath{{Figures/}}
\usepackage{caption}
\usepackage{subfig} 
\usepackage{amsmath}
\usepackage{amssymb}
\usepackage{epsfig}
\usepackage{cite}
\usepackage{color}
\usepackage{balance}
\usepackage{amsmath}
\usepackage{amsfonts} 
\usepackage{graphicx}
\usepackage{pdfpages}
\usepackage[T1]{fontenc} 
\usepackage{array}
\usepackage{url}

\usepackage{color}
\usepackage{wrapfig}
\usepackage{lipsum}
\usepackage[hidelinks]{hyperref}
\usepackage{multirow}
\usepackage{tabularx}
\usepackage{tikz}
\usepackage{nth} 
\usepackage{algpseudocode} 
\usepackage{algorithm,algpseudocode}
\usepackage{breqn}
\usepackage{todonotes}
\usepackage{verbatim}

\begin{document}
\title{Forecasting Precipitable Water Vapor Using LSTMs}
\author{\IEEEauthorblockN{
Mayank Jain\IEEEauthorrefmark{1},
Shilpa Manandhar\IEEEauthorrefmark{2},
Yee Hui Lee\IEEEauthorrefmark{2},
Stefan Winkler\IEEEauthorrefmark{3}, and
Soumyabrata Dev\IEEEauthorrefmark{1}\IEEEauthorrefmark{4}
}
\IEEEauthorblockA{\IEEEauthorrefmark{1} School of Computer Science, University College Dublin, Ireland}
\IEEEauthorblockA{\IEEEauthorrefmark{2} School of Electrical and Electronic Engineering, Nanyang Technological University (NTU), Singapore}
\IEEEauthorblockA{\IEEEauthorrefmark{3} School of Computing, National University of Singapore (NUS)}
\IEEEauthorblockA{\IEEEauthorrefmark{4} ADAPT SFI Research Centre, Dublin, Ireland}

\thanks{Send correspondence to M.\ Jain, E-mail: mayank.jain1@ucdconnect.ie.}
\vspace{-0.6cm}
}

\maketitle

\begin{abstract}
Long-Short-Term-Memory (LSTM) networks have been used extensively for time series forecasting in recent years due to their ability of learning patterns over different periods of time. In this paper, this ability is applied to learning the pattern of Global Positioning System (GPS)-based Precipitable Water Vapor (PWV) measurements over a period of $4$ hours. The trained model was evaluated on more than $1500$ hours of recorded data.  It achieves a root mean square error (RMSE) of $0.098$mm for a forecasting interval of $5$ minutes in the future, and outperforms the naive approach for a lead-time of up to $40$ minutes.
\end{abstract}

\IEEEpeerreviewmaketitle

\section{Introduction}
In recent years, GPS (Global Positioning System)-based PWV (Precipitable Water Vapor) values have proved very helpful in determining/forecasting rainfall events \cite{manandhar2019data,manandhar2018importance}. This has shifted the focus of forecasting from rainfall events to GPS-based PWV values. 

Long Short-Term Memory (LSTM) have shown their potential in  time series forecasting \cite{fischer2018deep}. Utilizing this potential, an LSTM-based Deep Neural Network (DNN) has been designed and trained in this paper\footnote{~The code is available at \url{https://github.com/jain15mayank/PWV-Forecasts-Using-LSTM}.} to successfully forecast GPS-based PWV values with high accuracy.

\section{GPS-based PWV Measurements}
\subsection{PWV Dataset and Pre-processing}

The PWV values are computed from GPS measurements in 5-minute intervals. The GPS signals are usually affected by two primary delays in the atmosphere -- Zenith Hydrostatic Delay ($ZHD$) and Zenith Wet Delay ($ZWD$). The $ZWD$ delay occurs owing to the water vapor content in the atmosphere. We compute PWV fom the $ZWD$ delays as follows:
\begin{dmath}
    PWV=PI\cdot ZWD
    \label{eq:PWV_ZWD}
\end{dmath}
\begin{dmath}
	PI=[-\textrm{sgn}(L_{a}) \cdot 1.7\cdot 10^{-5} |L_{a}|^{h_{fac}}-0.0001] \cdot \cos\frac{2\pi(DoY-28)}{365.25}+0.165-1.7\cdot 10^{-5}|L_{a}|^{1.65}+f,
    \label{eq:PI}
\end{dmath}
where $L_{a}$ refers to the latitude, $DoY$ is day-of-year, the value of $h_{fac}$ is $1.48$ for stations in northern hemisphere and $1.25$ for the southern hemisphere. We compute $f=-2.38\cdot 10^{-6}H$, where \textit{H} is the station height, and the $ZWD$ values are processed for a tropical IGS GPS station, ID: NTUS ($1.30$$^{\circ}$N, $103.68$$^{\circ}$E).

A windowed dataset is required for training the LSTM-based deep neural network for time-series. In this case, each window is a continuous slice of PWV measurements for $4$ hours straight (i.e.\ $48$ consecutive readings). The output label is the predicted value or the next consecutive reading in the dataset (i.e.\ $49^{th}$ consecutive reading following the values considered for the corresponding input window). The presence of multiple gaps (missing values in the raw data) has also been considered while pre-processing the dataset. This ultimately led to $90011$ windows of consecutive readings. In other words, this accounted for more than $7500$ hours of PWV measurement data. The first $80\%$ of this pre-processed dataset was used for training the network, while the remainder was used for testing and reporting results.

\subsection{Forecasting Methodology}

\begin{figure}[!ht]
    \centering
    \includegraphics[width=\columnwidth]{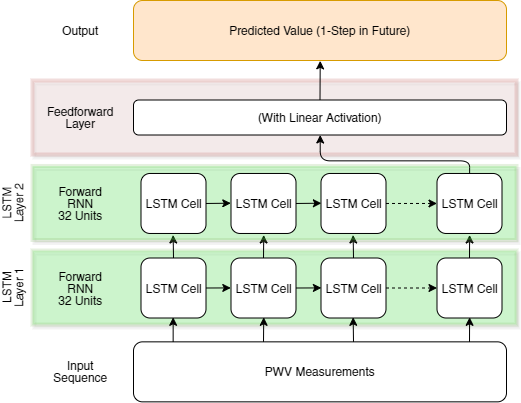}
    \caption{LSTM network model for PWV forecasting.}
    \label{fig:LSTMmodel}
    \vspace{-0.3cm}
\end{figure}

An LSTM-based deep neural network (see Fig.~\ref{fig:LSTMmodel}) has been trained for the task of predicting for a lead-time of $5$ minutes (i.e.\ immediate next step in series) given the past data of consecutive $4$ hours. Similar to the Recurrent Neural Network Language Model (RNNLM) \cite{Mikolov2011}, the trained network is used to forecast PWV values ahead into the future.

The model was trained with the Adam optimizer in \textit{Keras} using the default settings, but with a specially designed schedule for the learning rate $\eta$:
\begin{equation}
    \eta=\begin{cases}
            10^{-4} \times 10^{epoch/20} & \text{if } \eta<10^{-2},\\
            10^{-2} & \text{otherwise}.
        \end{cases}
    \label{eq:LRschedule}
\end{equation}
The schedule has been determined by running various experiments with varying learning rates in an attempt to minimize the loss. Further, for robust regression, Huber loss was used as the training metric \cite{huber1992robust}. The model was trained for $150$ epochs with a batch size of $32$ on the Google Colaboratory environment using GPU.

We observe that adding a constant bias of $-0.62$ to the trained model reduces its error rate considerably. This value is noted manually after training has been completed. The reason for adding this bias is because the last layer of the network is a simple feed-forward dense layer with $1$ neuron and linear activation. Hence, this bias is nothing but a minor modification in one of the network's weights itself.

\section{Results \& Discussions}
The trained DNN model is benchmarked against two popular baselines, which are used for time-series forecasting, namely, `average method' (where the average of considered past data is predicted as the future value) and `naive method' a.k.a.\ persistence (where the most recent past value is copied over as the predicted future value). The LSTM-based DNN model is noted to perform better than both the other baselines for a lead-time of up to $40$ minutes.

\begin{figure}[!ht]
    \centering
    \includegraphics[width=\columnwidth]{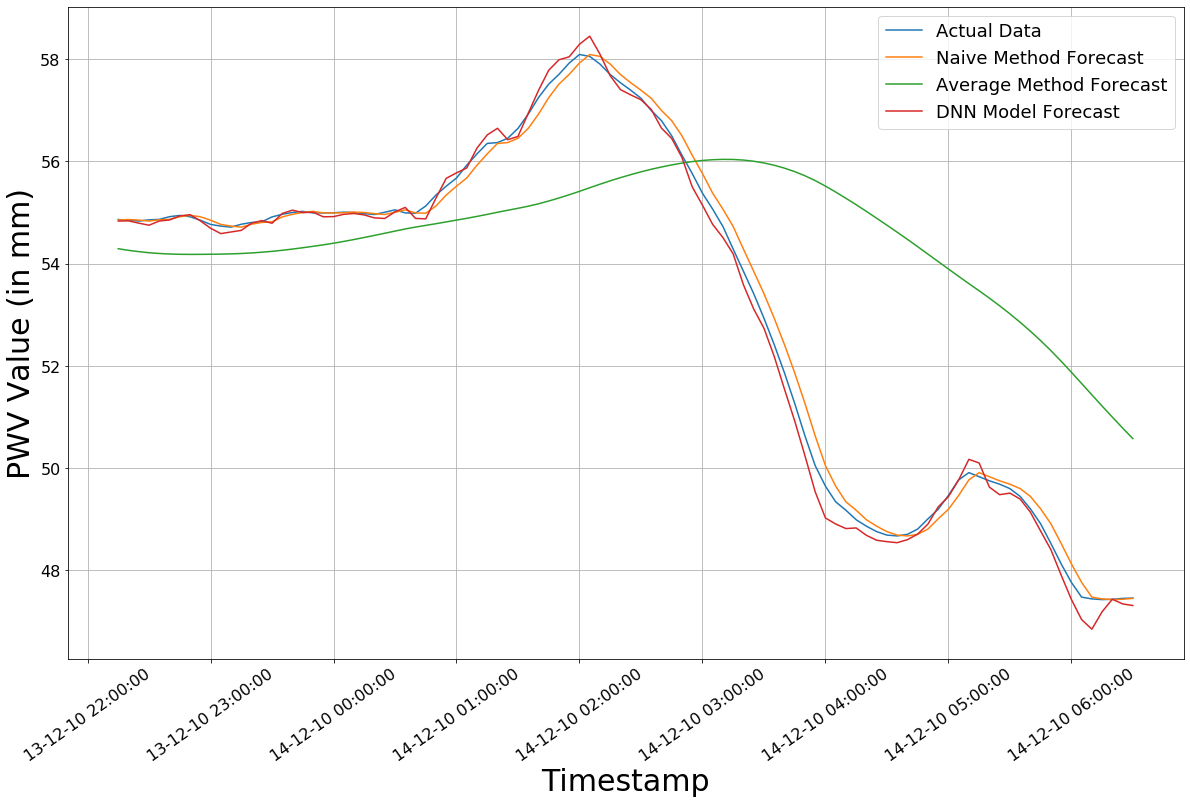}
    \caption{Comparison of DNN (LSTM) model predictions with baselines (for 15 minutes in future).}
    \label{fig:ModelComparison}
    \vspace{-0.3cm}
\end{figure}

From a qualitative perspective, the model captures the variations in the data fairly well. This  can be clearly seen in Figure \ref{fig:ModelComparison} which was generated by providing real data for $4$ hours before $15$ minutes of the plotted value.

\begin{figure}[!ht]
    \centering
    \includegraphics[width=\columnwidth]{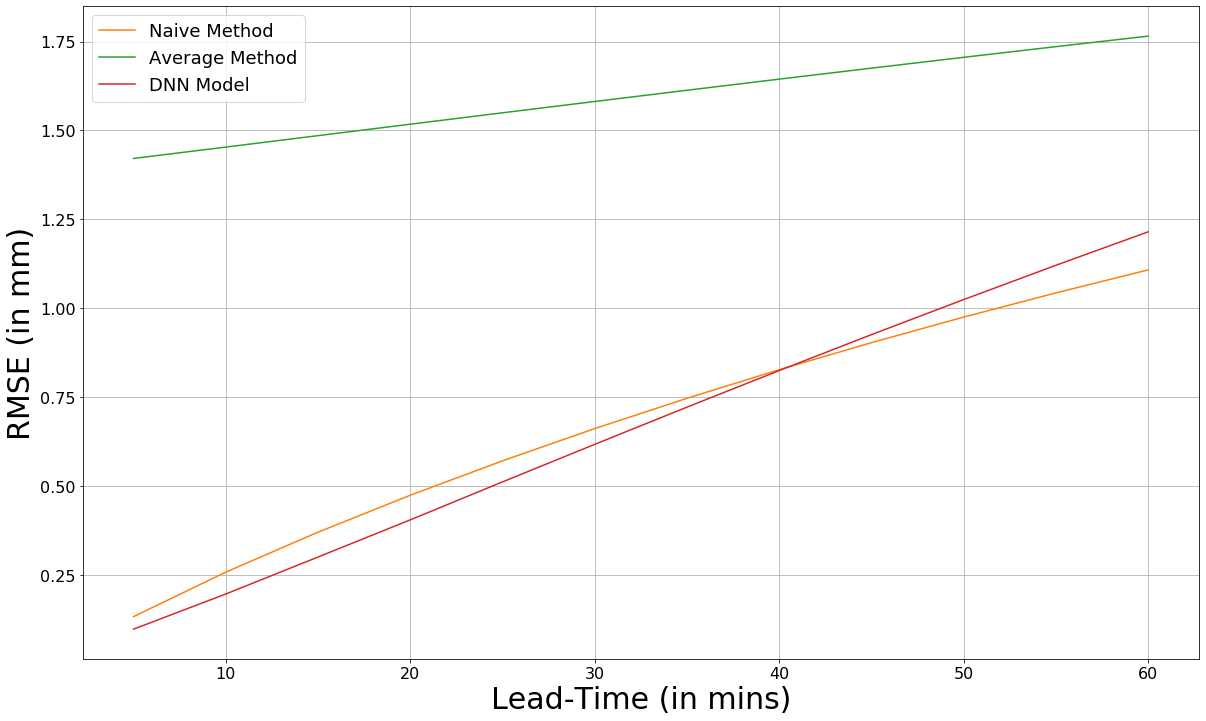}
    \caption{RMSE values for the DNN model and baselines over a range of lead-times (5-60 minutes).}
    \label{fig:RMSEComparison}
\end{figure}

To quantitatively analyze the results, Root Mean Square Error (RMSE) has been calculated over the complete test set for various lead-times. The results shown in Figure \ref{fig:RMSEComparison} demonstrate that the trained DNN model performs better than both baselines up to a lead-time of $40$ minutes. Moreover, with an increment in lead-time, the RMSE for DNN model also increases indicating that the error magnifies on each iteration. This is a likely possibility, as the future readings for larger lead-times were calculated using the approach of RNNLMs where the newly predicted value is assumed to be the actual value for future predictions.

Although, the performance of the trained DNN model is not very good for larger lead-times, the network demonstrates high accuracy at short-term forecasting. Table \ref{tab:RMSEvalues} shows the obtained RMSE values, averaged over the entire test set (more than $1500$ hours of recorded data), for varying lead times.

\begin{table}[!ht]
    \centering
    \caption{RMSE (mm) for different methods \& lead-times}
    \label{tab:RMSEvalues}
    \begin{tabular}{c c c c}
        \hline
        Lead-time & DNN Model & Naive Method & Average Method\\
        \hline
         5 min & \textrm{0.0978} & 0.1330 & 1.4212 \\
        10 min & \textrm{0.1966} & 0.2581 & 1.4532 \\
        15 min & \textrm{0.3005} & 0.3704 & 1.4854 \\
        \hline
    \end{tabular}
    \vspace{-0.2cm}
\end{table}

\section{Conclusion \& Future Work}
This paper presents an LSTM-based deep neural network for forecasting the future PWV values. We obtain good forecasting accuracy using our proposed framework as compared to other benchmarking methods. In the future, we intend to benchmark our LSTM-based network with other benchmarking methods \cite{manandhar2019predicting}, use longer time-period for statistical analysis, and include other sensor data \cite{IGARSS2015} for better prediction.


\bibliographystyle{IEEEbib}

\end{document}